\documentclass[aps,nofootinbib,prl,twocolumn,floatfix]{revtex4} %eqsecnum,
\usepackage{graphicx}
\input epsf
\newcommand{\sfig}[2]{
\includegraphics[width=#2]{#1}
        }
\newcommand{\Sfig}[2]{
    \begin{figure}[thbp]
    \sfig{#1.eps}{\columnwidth}
    \caption{{\small #2}}
    \label{fig:#1}
    \end{figure}
}
\newcommand{\Sfigs}[2]{
    \begin{figure}[thbp]
    \sfig{#1.eps}{0.9\columnwidth}
    \caption{{\small #2}}
    \label{fig:#1}
    \end{figure}
}

\newcommand{\rf}[1]{\ref{fig:#1}}

\def\cmm2{{\,\rm cm^{-2}}}
\def\cm2{{\,{\rm cm}^2}}
\def\cmm3{{\,{\rm cm}^{-3}}}
\def\gcmm3{{\,{\rm g\,cm^{-3}}}}

\def\fun#1#2{\lower3.6pt\vbox{\baselineskip0pt\lineskip.9pt
  \ialign{$\mathsurround=0pt#1\hfil##\hfil$\crcr#2\crcr\sim\crcr}}}

\def\be{\begin{equation}}
\def\ee{\end{equation}}
\def\bea{\begin{eqnarray}}
\def\eea{\end{eqnarray}}
\newcommand{\vs}{\nonumber\\}

\newcommand{\ec}[1]{Eq.\ (\ref{eq:#1})}

\newcommand{\eql}[1]{\label{eq:#1}}

\begin{document}
%\baselineskip=24pt
%\twocolumn[\hsize\textwidth\columnwidth\hsize\csname @twocolumnfalse\endcsname
%\pagestyle{empty}
%\begin{center}
%\rightline{{\large DRAFT} (Ewan; July 28, 1999)}
%\bigskip
%\rightline{FERMILAB--Pub--}
%\rightline{astro-ph/0002360}
%\rightline{submitted to {\it Phys. Rev. Lett.}}

%\vspace{.2in}
\title{Learning from the Scatter in Type Ia Supernovae }

%\vspace{.2in}
\author{Scott Dodelson$^{1,2}$ and  Alberto Vallinotto$^{1,3}$}
%\vspace{.2in}

\affiliation{$^1$Particle Astrophysics Center, Fermi National
Accelerator Laboratory, Batavia, IL~~60510-0500}
\affiliation{$^2$Department of Astronomy \& Astrophysics, The
University of Chicago, Chicago, IL~~60637-1433}
\affiliation{$^3$Department of Physics, The University of Chicago,
Chicago, IL~~60637-1433}

\date{\today}
%\smallskip
\begin{abstract}
Type Ia Supernovae are standard candles so their mean apparent
magnitude has been exploited to learn about the redshift-distance
relationship. Besides intrinsic scatter in this standard candle,
additional scatter is caused by gravitational magnification by 
large scale structure. Here we probe the dependence of this
 dispersion on cosmological parameters and show that
information about the amplitude of clustering, $\sigma_8$, is
contained in the scatter. In principle, it will be possible to
constrain $\sigma_8$ to within $5\%$ with observations of $2000$
Type Ia Supernovae. However, extracting this information requires
subtlety as the distribution of magnifications is far from Gaussian.
If one incorrectly assumes a Gaussian distribution, the estimate of the
clustering amplitude will be biased three-$\sigma$ away from the true value.
\end{abstract}
\maketitle

{\it Introduction.---} Type Ia Supernovae (SNIa) are standard
candles~\cite{baade}, with little dispersion around their mean
luminosity. By measuring their apparent magnitudes, therefore, we
can infer their distances from us. By observing supernovae at
cosmological distances, we can measure the redshift-distance
relationship and thereby extract information about cosmological
parameters~\cite{colgate,tammann,perl95}. Indeed, this method has
supplied the most direct argument to date for dark
energy~\cite{Riess:1998cb,Perlmutter:1998np} and serves as the basis
for future proposals to probe the nature of dark energy, such as the
Supernova Acceleration Probe (SNAP)~\cite{Aldering:2004ak}.

The success of this program is based on the small intrinsic scatter
in the SNIa luminosity. Various techniques have aided in the
reduction of this
scatter~\cite{phillips,1995ApJ...438L..17R,1996ApJ...473...88R,1997ApJ...483..565P},
which may be reduced even further in the
future~\cite{Aldering:2005qn}. However, the intrinsic dispersion of
SNIa luminosities is not the only source of scatter in the
observations. Images at cosmological distances can be magnified (or
de-magnified) by gravitational lensing produced by structure in the
universe~\cite{Kantowski:1995bd,Frieman:1996xk,1997ApJ...475L..81W,1998ApJ...507..483K}.
The amplitude of this {\it cosmic dispersion} depends on
cosmological parameters~\cite{Valageas:1999ch,1999MNRAS.305..746M}:
it increases with the matter density and the fluctuation amplitude.
In principle, then, it might be possible to extract information
about cosmological parameters not just by studying the mean apparent
magnitudes of SNIa but also by looking at the scatter around the
mean.

Since the mean is much more sensitive than the dispersion to the
matter density, little additional information
about $\Omega_m$ comes from the scatter. On the other hand, since the mean is
completely independent of the fluctuation amplitude, it may be
possible to use the cosmic dispersion profitably to infer
$\sigma_8$, the rms amplitude of fluctuations on a scale of
$8h^{-1}$Mpc. Ironically, in this age of precise parameter
determination from measuring fluctuations in the microwave
background and large scale structure, constraints on $\sigma_8$ are
very loose
%\footnote{The reason for this is that galaxy surveys probe
%the fluctuations multiplied by an unknown bias factor. Only the
%shape of the galaxy power spectrum is therefore used to constrain
%cosmological parameters, and the shape does not depend on
%$\sigma_8$. A similar, though more complicated, mechanism precludes
%strong constraints from the cosmic microwave background.}
. Current
estimates~\cite{Hoekstra:2002xs,Spergel:2003cb,Tegmark:2003ud,Seljak:2004xh,Viel:2004np,Sanchez:2005pi,Jarvis:2005ck,Viel:2005ha}
hover in the range $0.8-1.0$, and there is some evidence that
$\sigma_8$ may be even larger than
$1.0$~\cite{Komatsu:2002wc,Bond:2002tp} or as small as
$0.6$~\cite{Viana:2001dq}. This leads us to ask whether upcoming
supernovae searches can measure the cosmic dispersion and use it to
constrain $\sigma_8$.

{\it Distance Modulus.---} The distance modulus of an unlensed
source at redshift $z$ is \be \mu_0 =
5\log_{10}\left[\frac{d_L(z)}{10\, {\rm pc}}\right] \eql{defmu} \ee
where the luminosity distance in a flat universe (which is assumed
throughout) is \be d_L(z) = (1+z)c\int_0^z {dz'\over
H(z')}.\eql{defdl} \ee Here $H(z)$ is the Hubble expansion rate,
which in a flat universe with a cosmological constant and matter
takes the form $H(z)=H_0[\Omega_m(1+z)^3+(1-\Omega_m)]^{1/2}$, with
$\Omega_m$ the matter density in units of the critical density. The
current Hubble radius is $c/H_0=3000 h^{-1}$Mpc, and we set $h=0.72$
throughout.

The actual distance modulus $\mu$ of a SNIa at redshift $z$ differs
from that given in \ec{defmu} because of the \textit{intrinsic}
dispersion $\delta\mu_{\rm int}$ -- which is due to measurement
errors, dispersion in SNIa luminosities, and absorbtion
along the line of sight -- and because of the \textit{cosmic}
dispersion $\delta\mu_{\rm cos}$ due to gravitational lensing:
\begin{equation}
    \mu=\mu_0+\delta\mu_{\rm int}+\delta\mu_{\rm cos}.
\end{equation}
The mean of each of these two effects is zero, so
$\langle\mu\rangle=\mu_0$. The variance of $\delta\mu_{\rm cos}$
depends on the redshift of the source and can be related to the
variance of the \textit{convergence} $\kappa$
\begin{equation}\label{kappa}
    \kappa\equiv\frac{3\Omega_m H_0}{2\,c}\int_0^{\chi_s}d\chi
    [1+z(\chi)]\frac{\chi (\chi_s-\chi)}{\chi_s}\,\delta({\chi}),
\end{equation}
by the following integral along the line of sight over all Fourier
modes of the power spectrum: \bea \sigma_{\rm cos}^2 &=&
\left[\frac{5}{\ln(10)}\right]^2\langle\kappa^2\rangle\vs&=&{225\,\pi\Omega_m^2
H_0^4\over 4[\ln(10)]^2} \int_0^{\chi_s} d\chi
[1+z(\chi)]^2\frac{\chi^2 (\chi_s-\chi)^2}{\chi_s^2}\vs
 &&\times \int_0^\infty {dk\over k^2}
\Delta^2(k,z(\chi)). \eql{scos} \eea Here, $\chi$ is the comoving
distance to redshift $z$, $\chi_s\equiv \chi(z_s)$ denotes the
comoving distance to the source, $\delta(\chi)$ is the overdensity
at comoving distance $\chi$ and $\Delta^2=k^3P(k,z)/2\pi^2$ is a
dimensionless measure of the power.

% The subscript $_{\rm cos}$ here denotes that
%this is the cosmic scatter in the distance modulus to be
%distinguished from the internal dispersion due to non-uniformity of
%Type Ia supernovae or observational variations.

\Sfig{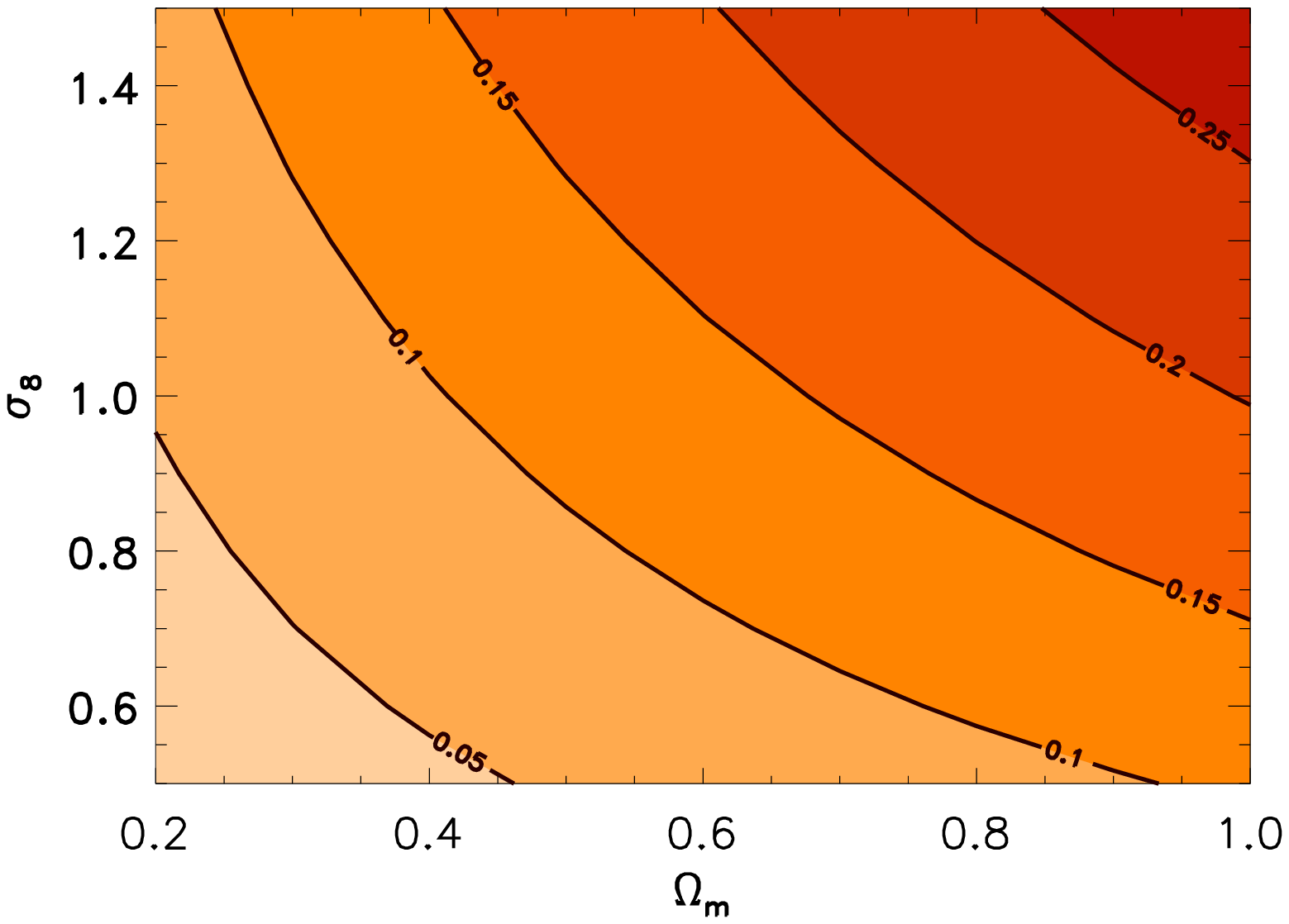}{Cosmic dispersion of SNIa distance modulus, $\sigma_{\rm cos}$,
as a function of cosmological parameters. Source here is at $z=1$.}

The integrand in \ec{scos} is here evaluated using the algorithm of
Smith et al.~\cite{Smith:2002dz} to generate the dark matter power
spectrum as a function of redshift. Fig.~\rf{delta} shows the cosmic
dispersion in the distance modulus as a function of the matter
density $\Omega_m$ and the fluctuation amplitude $\sigma_8$. Our results
agree with previous determinations~\cite{Hamana:1999rk,Valageas:1999ir}. The
curves of constant $\sigma_{\rm cos}$ have a familiar shape:
typically the amount of lensing increases if either the matter
density or the fluctuation amplitude goes up.

It is important to stress, however, that the cosmic
dispersion $\sigma_{\rm cos}$ does \textit{not} fully
characterize the distribution of magnifications and therefore of the
distance moduli. In particular, it has been shown
\cite{Valageas:1999ir} that the convergence -- and consequently also
the magnification -- is not distributed according to a Gaussian
distribution. As we show below, incorrect assumptions about the
underlying distribution can lead to a bias in $\sigma_8$.

{\it Projected Constraints on $\sigma_8$.---} The aim of the present
work is to extract information on the cosmological parameters not
only from the mean value of $\mu$ but also from its dispersion. This
operation is complicated by the fact that cosmic and intrinsic
dispersion add in quadrature, and therefore separating one from the
other requires some care~\cite{1999MNRAS.305..746M}. In what follows
the only assumption made about the intrinsic dispersion is that it
is independent of redshift. The conclusions presented below will
therefore be strengthened if
prior knowledge about the intrinsic dispersion can be used or
weakened if the intrinsic dispersion varies with redshift (unless this variation
is understood).

\Sfigs{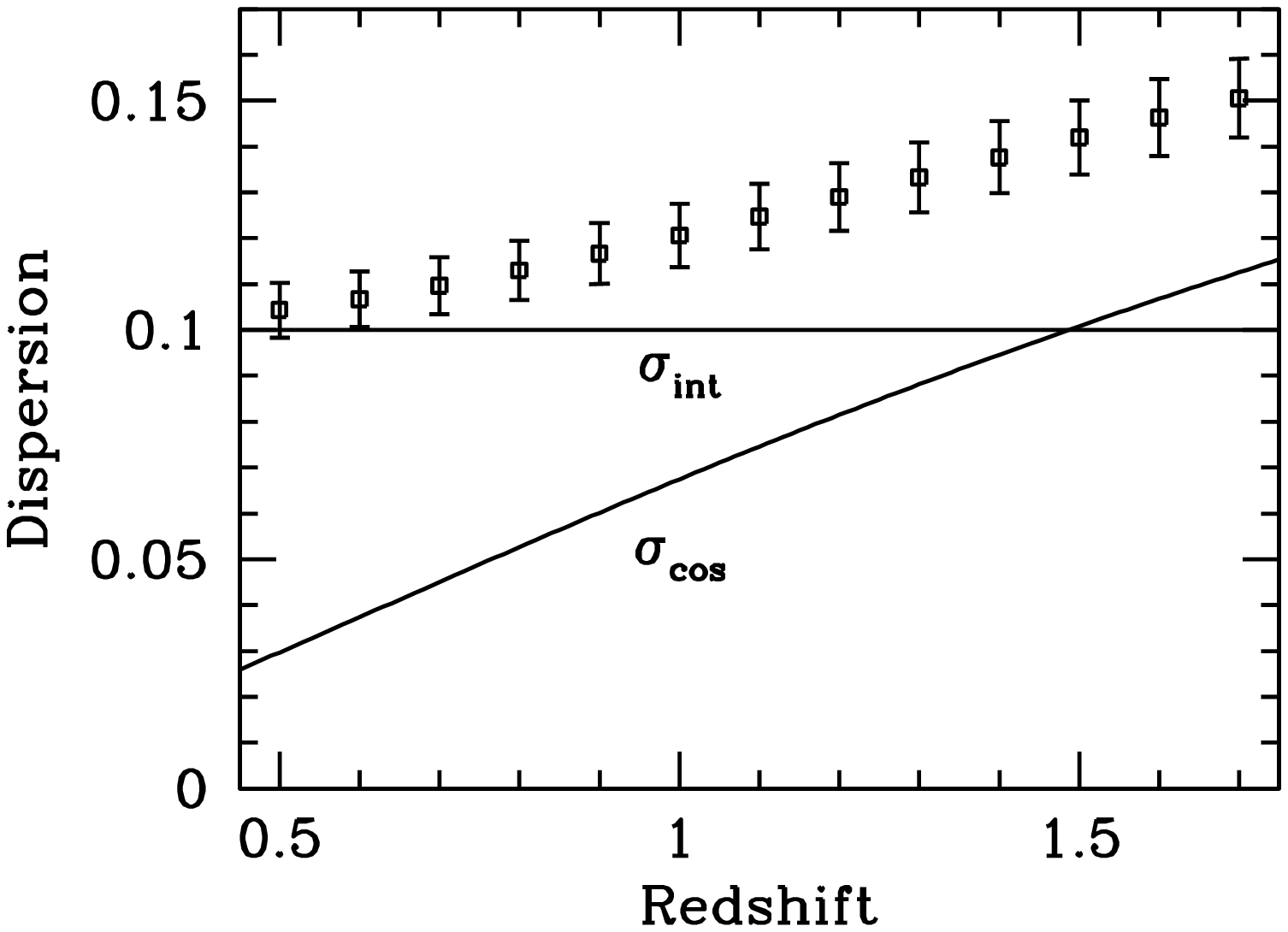}{Dispersion in distance modulus of a SNIa as a
function of its redshift for $\Omega_m=0.3$ and $\sigma_8=0.9$. The
two sources of dispersion -- internal and cosmic -- add in
quadrature. Throughout this work, we assume that the intrinsic
dispersion is constant and equal to $0.1$. Points with error bars
are projections for a future experiment which measures $2000$
supernovae in the redshift range $0.5$-$1.7$. Since the noise,
$\sigma_{\rm int}$ is unknown, the signal, $\sigma_{\rm cos}$, is
difficult to extract from the sum.}

Fig.~\rf{dispersion} shows the variation of the cosmic dispersion
with redshift. As expected, there is more lensing for more distant
sources, so the cosmic dispersion increases with $z$ (roughly as
$[1+z]^{2.5}$). This characteristic increase makes it possible to
distinguish cosmic dispersion from internal dispersion even without
any foreknowledge of the magnitude of the latter. The expected
errors on the total dispersion from a SNAP-like survey are also shown in
Fig.~\rf{dispersion}. The error on the total variance (the
dispersion squared) in a given redshift bin with $N$ supernovae
%\footnote{\textsc{Need to check the formula: is it
%$2\sigma^2/N$?.A}}
is $(2/N)^{1/2}\sigma^2$ or of order $10^{-3}$
for an intrinsic dispersion of $0.1$ (which will be assumed
throughout). However, to extract the cosmic dispersion, which is
expected to contribute of order half the total dispersion at high
redshifts, it is necessary to difference the variances in the
different redshift bins. This doubles the noise on the variance (for
two widely separated bins), and the signal is of order $\sigma_{\rm
cos}^2\sim (0.05)^2$, only marginally larger than the noise. A
careful weighting of the different redshift bins will then be
necessary to extract the signal.

{\it Likelihood analysis.---} If the likelihood function was known,
it would be straightforward to extract the cosmic signal optimally,
for the maximum of the likelihood is the minimal variance estimator.
A simple first guess for the likelihood is that $\delta\mu_{\rm
int}$ and the convergence $\kappa$ are Gaussian distributed with
variances $\sigma_{\rm int}^2$ and
$\langle\kappa^2\rangle=\sigma_{\kappa}^2$ respectively. (The
converegence $\kappa$ is considered because a fit for its
distribution, which will be used below, has been derived in Ref.~\cite{Wang:2002qc}.)
%A simple first guess
%for the likelihood is that the observed distance modulus for a
%supernova at redshift $z$ is distributed as a Gaussian with mean
%given by Eqs. (1) and (2) and variance equal to the sum of the
%cosmic variance and the intrisic variance: $C(z)=\sigma_{\rm
%cos}^2(z) + \sigma_{\rm int}^2$.
Assuming that the lensing is uncorrelated~\cite{Cooray:2005yr}, 
the total likelihood function is then the product of all the
Gaussian distributions corresponding to all the observed
supernovae. 
Here three parameters are considered:
$\sigma_8,\Omega_m$, and $\sigma_{\rm int}$. To project the errors
on these parameters, we carried out the following simulation:
\begin{itemize}

\item Generate a supernova redshift randomly chosen to lie in the interval\footnote{A more accurate range is $0.1-1.7$,
but this leads to problems when implementing the non-Gaussian distribution described below. We have checked
that the different ranges make little difference in the final projected errors.} $0.5$-$1.7$.

\item Using \ec{defmu}, compute the distance modulus of this SNIa
in a universe with $\Omega_m=0.3$ and $\sigma_8=0.9$.

\item With this set of cosmological parameters,
compute the cosmic dispersion using \ec{scos}.

\item Draw $\delta\mu_{\rm cos}$ from the distribution
resulting from the assumption that the underlying convergence is
distributed according to a Gaussian with mean zero and variance
$\sigma_{\kappa}^2$. Add $\delta\mu_{\rm cos}$ to the distance
modulus.

\item Draw $\delta\mu_{\rm in}$ from a Gaussian distribution with mean zero
and variance $\sigma_{\rm int}^2$. Add $\delta\mu_{\rm in}$ to the
distance modulus. This gives a final simulated value of $(z,\mu)$.
Repeat these steps $2000$ times.

\item For each point in the 3D parameter space
($\Omega_m,\sigma_8,\sigma_{\rm int}$), compute the likelihood of
getting these $2000$ data points.
% This likelihood function, as
%mentioned above, is assumed to be the product of $2000$ Gaussians.
\end{itemize}

Once the likelihood function has been obtained in the 3D parameter
space, projected errors on the cosmological parameters can be
calculated by marginalizing over the unknown $\sigma_{\rm int}$. The
resulting error matrix in $(\Omega_m,\sigma_8)$ space is roughly
diagonal: because the mean distance modulus determines $\Omega_m$
extremely accurately, the errors on $\sigma_8$ and $\Omega_m$ are
not correlated. That is, the mean distance modulus measurement
breaks the degeneracy in the dispersion shown in Fig.~\rf{delta}.
The projected 1-$\sigma$ error on $\sigma_8$ is $0.05$. This
projected error agrees well with that obtained via a Fisher matrix
analysis.

Up to this point, a Gaussian has been assumed for the distribution
of the convergence. However, the
distribution of the convergence is far from
Gaussian~\cite{Valageas:1999ir}: it is skewed, so that most
supernovae are demagnified while only a small fraction is highly
magnified. Wang et al.~\cite{Wang:2002qc} used N-Body simulations to
calibrate a phenomenological distribution based on the exact
theoretical results obtained in \cite{Valageas:1999ir,
Valageas:1999ch}. Assuming this calibrated fit as the correct
distribution for the convergence, we repeat the simulation with the following changes:
(i) Once $\sigma_{\kappa}^2(z)$ is computed, $\delta\mu_{\rm cos}$ is drawn
from the distribution obtained by assuming the results Wang et al.
for the convergence;
(ii) For each point in cosmological parameter space,
the likelihood function is determined by convolving the distribution
for $\delta\mu_{\rm cos}$ with a Gaussian of width $\sigma_{\rm
int}$; (iii)  To assess the impact that an erroneous
assumption would have on the analysis of experimental data, the
data generated with the non-Gaussian distribution are also
(incorrectly) analyzed assuming a Gaussian distribution for both the
convergence and the internal dispersion.

Fig.~\rf{wanggauss} shows the resulting likelihood function from one such simulation
in the $\sigma_8,\Omega_m$ plane after marginalizing over $\sigma_{\rm
int}$. The shaded contours show that
the maximum of the likelihood is shifted slightly from the
true value; this is a reasonable statistical fluctuation. The 1-$\sigma$ error
on $\sigma_8$ from this realization is $0.04$.
It is slightly smaller than that obtained if
the distribution were Gaussian. The shape of the distribution
function for the convergence therefore encodes even more information
about $\sigma_8$, information that could be mined if the
distribution function were known accurately enough.

\Sfig{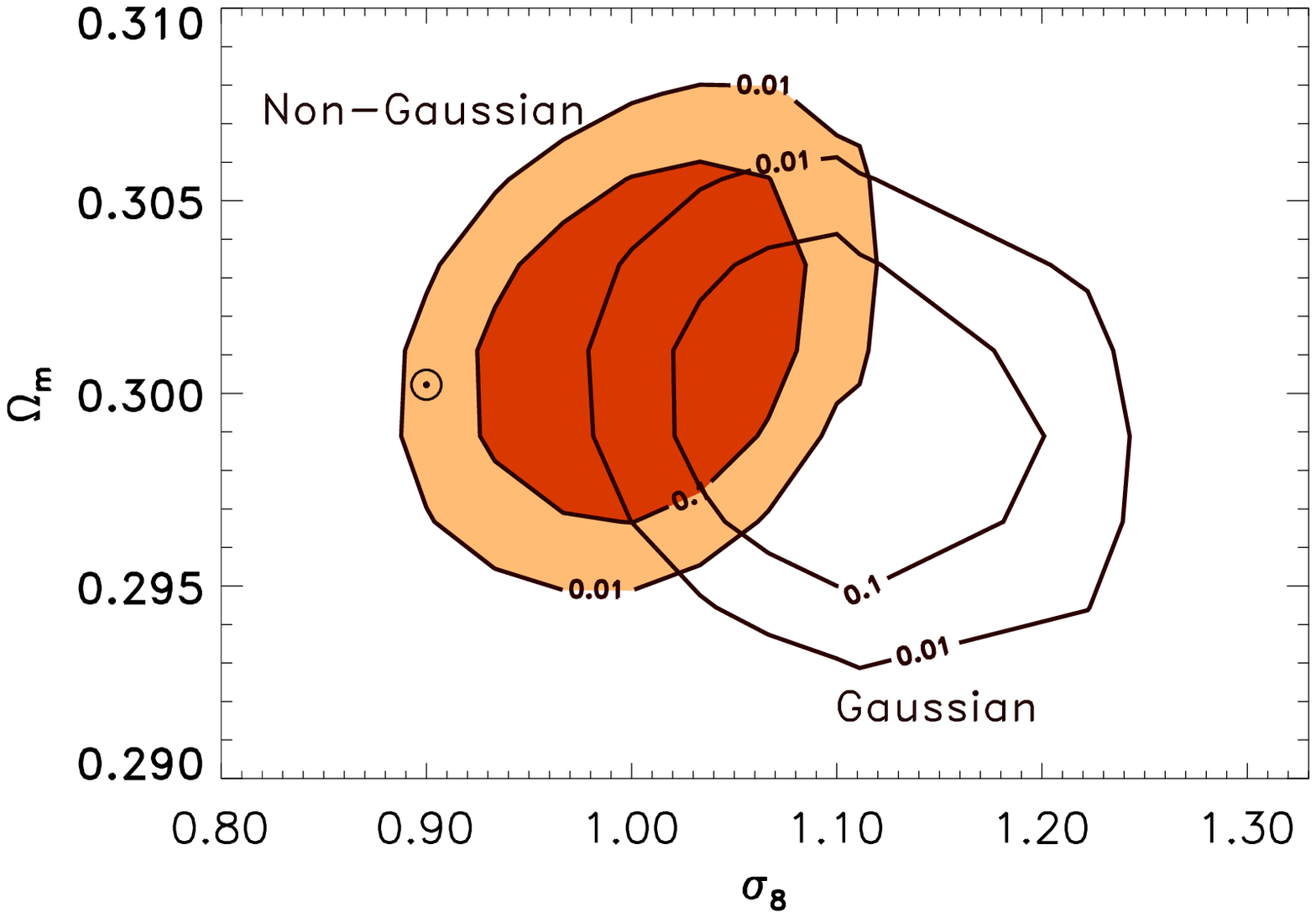}{Projected constraints on the matter density and
fluctuation amplitude from a SNAP-like survey. Contours show the
ratio of the likelihood to its maximum value (true value is at
starred point). The underlying probability distribution of the
convergence is non-Gaussian. The two sets of contours correspond to analyzing the simulted data
assuming the true distribution (shaded) or a Gaussian distribution (unshaded).}

However, with
the information encoded in this skewed distribution comes a
\textit{caveat}. An analysis which assumes that
the distribution is Gaussian will produce a biased estimate of $\sigma_8$. The unshaded contours
correspond to this assumption.
For this realization, the Gaussian assumption clearly leads to a worse estimate of
the best fit $\sigma_8$. To measure this bias, we ran one hundred simulations.
On average, the best fit $\sigma_8$ was equal to the true value if the non-Gaussian
analysis was used, but was biased with the Gaussian analysis.
The average bias was 
$\Delta\sigma_8= 0.12,$ three times larger than the anticipated
statistical error.
The skewness of the
distribution -- in particular, the few supernovae which have very
large magnifications -- can be explained in the Gaussian framework
only if the dispersion is very large. Therefore, the Gaussian
likelihood analysis will extract a value of $\sigma_8$
\textit{larger} than the true value.
The
lesson is: if we want to extract $\sigma_8$, or any clustering
parameter, from the dispersion of supernovae distance moduli, we
must account for the non-Gaussianity of the lensing distribution.
Note that this conclusion does {\it not} conflict with the recent
results of Ref.~\cite{Holz:2004xx}.  They showed that the
non-Gaussianity of the distribution does not bias the extraction of
the matter density or dark energy equation of state. We verify that
the matter density, which is largely determined by the mean distance
modulus, is unbiased in our simulations \textit{even} if the
Gaussian likelihood is used to analyze data generated with the
non-Gaussian distribution. The clustering parameter, $\sigma_8$,
though is biased because it is solely determined by the dispersion.

{\it Conclusions.---} Future SNIa surveys will be able to constrain
the clustering amplitude $\sigma_8$ to within $5\%$. This is
significantly better than current efforts and likely to be
competitive even with future measurements. In order to extract an
accurate value of $\sigma_8$, careful theoretical studies will need
to pin down not just the cosmic dispersion as a function of
cosmological parameters, but also the distribution of magnifications
(especially at low redshift).

Besides the bias that can be induced by neglecting the
non-Gaussianity of the convergence distribution, there are a number
of systematics that could complicate this determination. First, the
internal dispersion may vary with redshift: if this variation cannot
be understood, at least some prior knowledge of the magnitude of the
internal dispersion will be needed. Second, one might question the
$\sigma_8-\sigma_{\rm cos}$ relationship. If only dark matter
determined the lensing, then this connection would be relatively
straightforward. However, cosmic dispersion is determined by
structure on small scales. On the smallest scales,  one must worry
about the impact of baryons. Several groups
~\cite{White:2004kv,Zhan:2004wq} have studied this recently in a
different context and suggested that the power spectrum will be
affected on scales $k>k_f=30$h Mpc$^{-1}$. If so, this would bias
the determination of $\sigma_8$ at the ten percent level.
Hydrodynamical simulations, though, should be able to reduce this
systematic.

\acknowledgments{}

It is a pleasure to thank Andrea Mignone for the extensive
discussion of the numerical aspects in the early stages of the
project. This work is supported by the DOE, by NASA grant
NAG5-10842.

\bibliography{v5}

\begin{thebibliography}{37}
\expandafter\ifx\csname natexlab\endcsname\relax\def\natexlab#1{#1}\fi
\expandafter\ifx\csname bibnamefont\endcsname\relax
  \def\bibnamefont#1{#1}\fi
\expandafter\ifx\csname bibfnamefont\endcsname\relax
  \def\bibfnamefont#1{#1}\fi
\expandafter\ifx\csname citenamefont\endcsname\relax
  \def\citenamefont#1{#1}\fi
\expandafter\ifx\csname url\endcsname\relax
  \def\url#1{\texttt{#1}}\fi
\expandafter\ifx\csname urlprefix\endcsname\relax\def\urlprefix{URL }\fi
\providecommand{\bibinfo}[2]{#2}
\providecommand{\eprint}[2][]{\url{#2}}

\bibitem[{\citenamefont{{Baade}}(1938)}]{baade}
\bibinfo{author}{\bibfnamefont{W.}~\bibnamefont{{Baade}}},
  \bibinfo{journal}{\apj} \textbf{\bibinfo{volume}{88}}, \bibinfo{pages}{285}
  (\bibinfo{year}{1938}).

\bibitem[{\citenamefont{{Colgate}}(1979)}]{colgate}
\bibinfo{author}{\bibfnamefont{S.~A.} \bibnamefont{{Colgate}}},
  \bibinfo{journal}{\apj} \textbf{\bibinfo{volume}{232}}, \bibinfo{pages}{404}
  (\bibinfo{year}{1979}).

\bibitem[{\citenamefont{{Tammann}}(1979)}]{tammann}
\bibinfo{author}{\bibfnamefont{G.~A.} \bibnamefont{{Tammann}}}, in
  \emph{\bibinfo{booktitle}{Scientific Research with the Space Telescope}}
  (\bibinfo{year}{1979}), pp. \bibinfo{pages}{263--293}.

\bibitem[{\citenamefont{{Goobar} and {Perlmutter}}(1995)}]{perl95}
\bibinfo{author}{\bibfnamefont{A.}~\bibnamefont{{Goobar}}} \bibnamefont{and}
  \bibinfo{author}{\bibfnamefont{S.}~\bibnamefont{{Perlmutter}}},
  \bibinfo{journal}{\apj} \textbf{\bibinfo{volume}{450}}, \bibinfo{pages}{14}
  (\bibinfo{year}{1995}).

\bibitem[{\citenamefont{Riess et~al.}(1998)}]{Riess:1998cb}
\bibinfo{author}{\bibfnamefont{A.~G.} \bibnamefont{Riess}} \bibnamefont{et~al.}
  (\bibinfo{collaboration}{Supernova Search Team}), \bibinfo{journal}{Astron.
  J.} \textbf{\bibinfo{volume}{116}}, \bibinfo{pages}{1009}
  (\bibinfo{year}{1998}), \eprint{astro-ph/9805201}.

\bibitem[{\citenamefont{Perlmutter et~al.}(1999)}]{Perlmutter:1998np}
\bibinfo{author}{\bibfnamefont{S.}~\bibnamefont{Perlmutter}}
  \bibnamefont{et~al.} (\bibinfo{collaboration}{Supernova Cosmology Project}),
  \bibinfo{journal}{Astrophys. J.} \textbf{\bibinfo{volume}{517}},
  \bibinfo{pages}{565} (\bibinfo{year}{1999}), \eprint{astro-ph/9812133}.

\bibitem[{\citenamefont{Aldering et~al.}(2004)}]{Aldering:2004ak}
\bibinfo{author}{\bibfnamefont{G.}~\bibnamefont{Aldering}} \bibnamefont{et~al.}
  (\bibinfo{collaboration}{SNAP}) (\bibinfo{year}{2004}),
  \eprint{astro-ph/0405232}.

\bibitem[{\citenamefont{{Phillips}}(1993)}]{phillips}
\bibinfo{author}{\bibfnamefont{M.~M.} \bibnamefont{{Phillips}}},
  \bibinfo{journal}{\apj} \textbf{\bibinfo{volume}{413}}, \bibinfo{pages}{L105}
  (\bibinfo{year}{1993}).

\bibitem[{\citenamefont{{Riess} et~al.}(1995)\citenamefont{{Riess}, {Press},
  and {Kirshner}}}]{1995ApJ...438L..17R}
\bibinfo{author}{\bibfnamefont{A.~G.} \bibnamefont{{Riess}}},
  \bibinfo{author}{\bibfnamefont{W.~H.} \bibnamefont{{Press}}},
  \bibnamefont{and} \bibinfo{author}{\bibfnamefont{R.~P.}
  \bibnamefont{{Kirshner}}}, \bibinfo{journal}{\apj}
  \textbf{\bibinfo{volume}{438}}, \bibinfo{pages}{L17} (\bibinfo{year}{1995}).

\bibitem[{\citenamefont{{Riess} et~al.}(1996)\citenamefont{{Riess}, {Press},
  and {Kirshner}}}]{1996ApJ...473...88R}
\bibinfo{author}{\bibfnamefont{A.~G.} \bibnamefont{{Riess}}},
  \bibinfo{author}{\bibfnamefont{W.~H.} \bibnamefont{{Press}}},
  \bibnamefont{and} \bibinfo{author}{\bibfnamefont{R.~P.}
  \bibnamefont{{Kirshner}}}, \bibinfo{journal}{\apj}
  \textbf{\bibinfo{volume}{473}}, \bibinfo{pages}{88} (\bibinfo{year}{1996}).

\bibitem[{\citenamefont{{Perlmutter} et~al.}(1997)\citenamefont{{Perlmutter},
  {Gabi}, {Goldhaber}, {Goobar}, {Groom}, {Hook}, {Kim}, {Kim}, {Lee}, {Pain}
  et~al.}}]{1997ApJ...483..565P}
\bibinfo{author}{\bibfnamefont{S.}~\bibnamefont{{Perlmutter}}},
  \bibinfo{author}{\bibfnamefont{S.}~\bibnamefont{{Gabi}}},
  \bibinfo{author}{\bibfnamefont{G.}~\bibnamefont{{Goldhaber}}},
  \bibinfo{author}{\bibfnamefont{A.}~\bibnamefont{{Goobar}}},
  \bibinfo{author}{\bibfnamefont{D.~E.} \bibnamefont{{Groom}}},
  \bibinfo{author}{\bibfnamefont{I.~M.} \bibnamefont{{Hook}}},
  \bibinfo{author}{\bibfnamefont{A.~G.} \bibnamefont{{Kim}}},
  \bibinfo{author}{\bibfnamefont{M.~Y.} \bibnamefont{{Kim}}},
  \bibinfo{author}{\bibfnamefont{J.~C.} \bibnamefont{{Lee}}},
  \bibinfo{author}{\bibfnamefont{R.}~\bibnamefont{{Pain}}},
  \bibnamefont{et~al.}, \bibinfo{journal}{\apj} \textbf{\bibinfo{volume}{483}},
  \bibinfo{pages}{565} (\bibinfo{year}{1997}).

\bibitem[{\citenamefont{Aldering}(2005)}]{Aldering:2005qn}
\bibinfo{author}{\bibfnamefont{G.}~\bibnamefont{Aldering}}
  (\bibinfo{year}{2005}), \eprint{astro-ph/0507426}.

\bibitem[{\citenamefont{Kantowski et~al.}(1995)\citenamefont{Kantowski,
  Vaughan, and Branch}}]{Kantowski:1995bd}
\bibinfo{author}{\bibfnamefont{R.}~\bibnamefont{Kantowski}},
  \bibinfo{author}{\bibfnamefont{T.}~\bibnamefont{Vaughan}}, \bibnamefont{and}
  \bibinfo{author}{\bibfnamefont{D.}~\bibnamefont{Branch}},
  \bibinfo{journal}{Astrophys. J.} \textbf{\bibinfo{volume}{447}},
  \bibinfo{pages}{35} (\bibinfo{year}{1995}), \eprint{astro-ph/9511108}.

\bibitem[{\citenamefont{Frieman}(1997)}]{Frieman:1996xk}
\bibinfo{author}{\bibfnamefont{J.~A.} \bibnamefont{Frieman}},
  \bibinfo{journal}{Comments Astrophys.} \textbf{\bibinfo{volume}{18}},
  \bibinfo{pages}{323} (\bibinfo{year}{1997}), \eprint{astro-ph/9608068}.

\bibitem[{\citenamefont{{Wambsganss} et~al.}(1997)\citenamefont{{Wambsganss},
  {Cen}, {Xu}, and {Ostriker}}}]{1997ApJ...475L..81W}
\bibinfo{author}{\bibfnamefont{J.}~\bibnamefont{{Wambsganss}}},
  \bibinfo{author}{\bibfnamefont{R.}~\bibnamefont{{Cen}}},
  \bibinfo{author}{\bibfnamefont{G.}~\bibnamefont{{Xu}}}, \bibnamefont{and}
  \bibinfo{author}{\bibfnamefont{J.~P.} \bibnamefont{{Ostriker}}},
  \bibinfo{journal}{\apj} \textbf{\bibinfo{volume}{475}}, \bibinfo{pages}{L81+}
  (\bibinfo{year}{1997}).

\bibitem[{\citenamefont{{Kantowski}}(1998)}]{1998ApJ...507..483K}
\bibinfo{author}{\bibfnamefont{R.}~\bibnamefont{{Kantowski}}},
  \bibinfo{journal}{\apj} \textbf{\bibinfo{volume}{507}}, \bibinfo{pages}{483}
  (\bibinfo{year}{1998}).

\bibitem[{\citenamefont{Valageas}(2000)}]{Valageas:1999ch}
\bibinfo{author}{\bibfnamefont{P.}~\bibnamefont{Valageas}},
  \bibinfo{journal}{Astron. Astrophys.} \textbf{\bibinfo{volume}{354}},
  \bibinfo{pages}{767} (\bibinfo{year}{2000}), \eprint{astro-ph/9904300}.

\bibitem[{\citenamefont{{Metcalf}}(1999)}]{1999MNRAS.305..746M}
\bibinfo{author}{\bibfnamefont{R.~B.} \bibnamefont{{Metcalf}}},
  \bibinfo{journal}{Mon. Not. Roy. Astron. Soc.}
  \textbf{\bibinfo{volume}{305}}, \bibinfo{pages}{746} (\bibinfo{year}{1999}).

\bibitem[{\citenamefont{Hoekstra et~al.}(2002)\citenamefont{Hoekstra, Yee, and
  Gladders}}]{Hoekstra:2002xs}
\bibinfo{author}{\bibfnamefont{H.}~\bibnamefont{Hoekstra}},
  \bibinfo{author}{\bibfnamefont{H.~K.~C.} \bibnamefont{Yee}},
  \bibnamefont{and} \bibinfo{author}{\bibfnamefont{M.~D.}
  \bibnamefont{Gladders}}, \bibinfo{journal}{Astrophys. J.}
  \textbf{\bibinfo{volume}{577}}, \bibinfo{pages}{595} (\bibinfo{year}{2002}),
  \eprint{astro-ph/0204295}.

\bibitem[{\citenamefont{Spergel et~al.}(2003)}]{Spergel:2003cb}
\bibinfo{author}{\bibfnamefont{D.~N.} \bibnamefont{Spergel}}
  \bibnamefont{et~al.} (\bibinfo{year}{2003}), \eprint{astro-ph/0302209}.

\bibitem[{\citenamefont{Tegmark et~al.}(2004)}]{Tegmark:2003ud}
\bibinfo{author}{\bibfnamefont{M.}~\bibnamefont{Tegmark}} \bibnamefont{et~al.}
  (\bibinfo{collaboration}{SDSS}), \bibinfo{journal}{Phys. Rev.}
  \textbf{\bibinfo{volume}{D69}}, \bibinfo{pages}{103501}
  (\bibinfo{year}{2004}), \eprint{astro-ph/0310723}.

\bibitem[{\citenamefont{Seljak et~al.}(2005)}]{Seljak:2004xh}
\bibinfo{author}{\bibfnamefont{U.}~\bibnamefont{Seljak}} \bibnamefont{et~al.},
  \bibinfo{journal}{Phys. Rev.} \textbf{\bibinfo{volume}{D71}},
  \bibinfo{pages}{103515} (\bibinfo{year}{2005}), \eprint{astro-ph/0407372}.

\bibitem[{\citenamefont{Viel et~al.}(2004)\citenamefont{Viel, Weller, and
  Haehnelt}}]{Viel:2004np}
\bibinfo{author}{\bibfnamefont{M.}~\bibnamefont{Viel}},
  \bibinfo{author}{\bibfnamefont{J.}~\bibnamefont{Weller}}, \bibnamefont{and}
  \bibinfo{author}{\bibfnamefont{M.}~\bibnamefont{Haehnelt}},
  \bibinfo{journal}{Mon. Not. Roy. Astron. Soc.}
  \textbf{\bibinfo{volume}{355}}, \bibinfo{pages}{L23} (\bibinfo{year}{2004}),
  \eprint{astro-ph/0407294}.

\bibitem[{\citenamefont{Sanchez et~al.}(2005)}]{Sanchez:2005pi}
\bibinfo{author}{\bibfnamefont{A.~G.} \bibnamefont{Sanchez}}
  \bibnamefont{et~al.} (\bibinfo{year}{2005}), \eprint{astro-ph/0507583}.

\bibitem[{\citenamefont{Jarvis et~al.}(2005)\citenamefont{Jarvis, Jain,
  Bernstein, and Dolney}}]{Jarvis:2005ck}
\bibinfo{author}{\bibfnamefont{M.}~\bibnamefont{Jarvis}},
  \bibinfo{author}{\bibfnamefont{B.}~\bibnamefont{Jain}},
  \bibinfo{author}{\bibfnamefont{G.}~\bibnamefont{Bernstein}},
  \bibnamefont{and} \bibinfo{author}{\bibfnamefont{D.}~\bibnamefont{Dolney}}
  (\bibinfo{year}{2005}), \eprint{astro-ph/0502243}.

\bibitem[{\citenamefont{Viel and Haehnelt}(2005)}]{Viel:2005ha}
\bibinfo{author}{\bibfnamefont{M.}~\bibnamefont{Viel}} \bibnamefont{and}
  \bibinfo{author}{\bibfnamefont{M.~G.} \bibnamefont{Haehnelt}}
  (\bibinfo{year}{2005}), \eprint{astro-ph/0508177}.

\bibitem[{\citenamefont{Komatsu and Seljak}(2002)}]{Komatsu:2002wc}
\bibinfo{author}{\bibfnamefont{E.}~\bibnamefont{Komatsu}} \bibnamefont{and}
  \bibinfo{author}{\bibfnamefont{U.}~\bibnamefont{Seljak}},
  \bibinfo{journal}{Mon. Not. Roy. Astron. Soc.}
  \textbf{\bibinfo{volume}{336}}, \bibinfo{pages}{1256} (\bibinfo{year}{2002}),
  \eprint{astro-ph/0205468}.

\bibitem[{\citenamefont{Bond et~al.}(2005)}]{Bond:2002tp}
\bibinfo{author}{\bibfnamefont{J.~R.} \bibnamefont{Bond}} \bibnamefont{et~al.},
  \bibinfo{journal}{Astrophys. J.} \textbf{\bibinfo{volume}{626}},
  \bibinfo{pages}{12} (\bibinfo{year}{2005}), \eprint{astro-ph/0205386}.

\bibitem[{\citenamefont{Viana et~al.}(2002)\citenamefont{Viana, Nichol, and
  Liddle}}]{Viana:2001dq}
\bibinfo{author}{\bibfnamefont{P.~T.~P.} \bibnamefont{Viana}},
  \bibinfo{author}{\bibfnamefont{R.~C.} \bibnamefont{Nichol}},
  \bibnamefont{and} \bibinfo{author}{\bibfnamefont{A.~R.}
  \bibnamefont{Liddle}}, \bibinfo{journal}{Astrophys. J.}
  \textbf{\bibinfo{volume}{569}}, \bibinfo{pages}{L75} (\bibinfo{year}{2002}),
  \eprint{astro-ph/0111394}.

\bibitem[{\citenamefont{Smith et~al.}(2002)}]{Smith:2002dz}
\bibinfo{author}{\bibfnamefont{R.~E.} \bibnamefont{Smith}} \bibnamefont{et~al.}
  (\bibinfo{collaboration}{The Virgo Consortium}) (\bibinfo{year}{2002}),
  \eprint{astro-ph/0207664}.

\bibitem[{\citenamefont{Hamana and Futamase}(1999)}]{Hamana:1999rk}
\bibinfo{author}{\bibfnamefont{T.}~\bibnamefont{Hamana}} \bibnamefont{and}
  \bibinfo{author}{\bibfnamefont{T.}~\bibnamefont{Futamase}}
  (\bibinfo{year}{1999}), \eprint{astro-ph/9912319}.

\bibitem[{\citenamefont{Valageas}(1999)}]{Valageas:1999ir}
\bibinfo{author}{\bibfnamefont{P.}~\bibnamefont{Valageas}}
  (\bibinfo{year}{1999}), \eprint{astro-ph/9911336}.

\bibitem[{\citenamefont{Wang et~al.}(2002)\citenamefont{Wang, Holz, and
  Munshi}}]{Wang:2002qc}
\bibinfo{author}{\bibfnamefont{Y.}~\bibnamefont{Wang}},
  \bibinfo{author}{\bibfnamefont{D.~E.} \bibnamefont{Holz}}, \bibnamefont{and}
  \bibinfo{author}{\bibfnamefont{D.}~\bibnamefont{Munshi}},
  \bibinfo{journal}{Astrophys. J.} \textbf{\bibinfo{volume}{572}},
  \bibinfo{pages}{L15} (\bibinfo{year}{2002}), \eprint{astro-ph/0204169}.

\bibitem[{\citenamefont{Cooray et~al.}(2005)\citenamefont{Cooray, Huterer, and
  Holz}}]{Cooray:2005yr}
\bibinfo{author}{\bibfnamefont{A.}~\bibnamefont{Cooray}},
  \bibinfo{author}{\bibfnamefont{D.}~\bibnamefont{Huterer}}, \bibnamefont{and}
  \bibinfo{author}{\bibfnamefont{D.}~\bibnamefont{Holz}}
  (\bibinfo{year}{2005}), \eprint{astro-ph/0509581}.

\bibitem[{\citenamefont{Holz and Linder}(2004)}]{Holz:2004xx}
\bibinfo{author}{\bibfnamefont{D.~E.} \bibnamefont{Holz}} \bibnamefont{and}
  \bibinfo{author}{\bibfnamefont{E.~V.} \bibnamefont{Linder}}
  (\bibinfo{year}{2004}), \eprint{astro-ph/0412173}.

\bibitem[{\citenamefont{White}(2004)}]{White:2004kv}
\bibinfo{author}{\bibfnamefont{M.~J.} \bibnamefont{White}},
  \bibinfo{journal}{Astropart. Phys.} \textbf{\bibinfo{volume}{22}},
  \bibinfo{pages}{211} (\bibinfo{year}{2004}), \eprint{astro-ph/0405593}.

\bibitem[{\citenamefont{Zhan and Knox}(2004)}]{Zhan:2004wq}
\bibinfo{author}{\bibfnamefont{H.}~\bibnamefont{Zhan}} \bibnamefont{and}
  \bibinfo{author}{\bibfnamefont{L.}~\bibnamefont{Knox}},
  \bibinfo{journal}{Astrophys. J.} \textbf{\bibinfo{volume}{616}},
  \bibinfo{pages}{L75} (\bibinfo{year}{2004}), \eprint{astro-ph/0409198}.

\end{thebibliography}
\end{document}